\documentclass[useAMS,usenatbib,A4]{mn2e}
\usepackage{graphicx}
\usepackage{amsmath}
\usepackage{amssymb,latexsym,amsopn}
\usepackage{lscape}
\title[The observed properties of FRBs]{The observed properties of Fast Radio Bursts}

\author[Vikram Ravi]{Vikram Ravi$^{1}$\thanks{E-mail: vikram@caltech.edu} 
\\
$^{1}$Cahill Center for Astronomy and Astrophysics, MC 249-17, California Institute of Technology, Pasadena, CA 91125, USA.\\
}

\begin{document}

\date{}

\pagerange{1--17} \pubyear{2017}

\maketitle

\label{firstpage}

\begin{abstract}

I present an empirical study of the properties of fast radio bursts (FRBs): Gigahertz-frequency, dispersed pulses of 
extragalactic origin. I focus my investigation on the sample of seventeen FRBs detected at the Parkes radio telescope with 
largely self-consistent instrumentation. Of this sample, six are temporally unresolved, eight exhibit evidence 
for scattering in inhomogeneous plasma, and five display potentially intrinsic temporal structure. 
The characteristic scattering timescales at a frequency of 1\,GHz range between 0.005\,ms and 32\,ms; moderate evidence 
exists for a relation between FRB scattering timescales and dispersion measures. Additionally, I present constraints on the 
fluences of Parkes FRBs, accounting for their uncertain sky-positions, and use the multiple-beam detection of FRB\,010724 (the Lorimer burst)
to measure its fluence to be $800\pm400$\,Jy\,ms. FRBs, including the repeating FRB\,121102, appear to manifest with a plethora 
of characteristics, and it is uncertain at present whether they share a common class of progenitor object, or arise from 
a selection of independent progenitors.

\end{abstract}


\section{Introduction}

A fast radio burst (FRB) may be broadly defined  \citep[cf.][]{pbb+15,kp15} as a demonstrably astrophysical, dispersed radio pulse, 
with a dispersion measure (DM) that significantly exceeds any estimate \citep[e.g.,][]{cl02} of the Milky Way free-electron column 
density along its sightline. Implicit in this definition is the condition that FRBs adhere to the cold, sparse plasma dispersion law 
\citep[e.g.,][]{k16}. The identification of the host galaxy of the repeating FRB\,121102 \citep{clw+17} has confirmed the existence 
of extragalactic sources of pulsed radio emission. Questions regarding the origins of FRBs may therefore be framed in 
terms of what kinds of sources, at what extragalactic distances, produce FRBs. 

Twenty-four FRB detections have now been published.\footnote{A catalogue of FRBs is maintained at 
http://www.astronomy.swin.edu.au/pulsar/frbcat/ \citep{pbj+16}.} Even among this small sample, the diversity 
of FRB properties is striking. FRB\,121102, the sole detection at the Arecibo Observatory, is also the only known repeater. 
Additionally, different FRBs display markedly different propagation signatures; for example, 
the timescales by which FRBs are temporally broadened during propagation by scattering due to plasma-density inhomogeneities vary 
by four orders of magnitude \citep{rsb+16}. Third, as I shall show, even among the sample of seventeen FRBs detected 
at the Parkes telescope, the range of fluences spans three orders of magnitude.

The primary goal of this paper is to homogenise the inference of FRB properties among the Parkes sample (\S2). In particular, I focus on 
self-consistent estimates of FRB fluences, DMs, intrinsic widths, and scattering timescales and frequency-dependencies. 
Such measurements have been compiled (and performed) by \citet{pbj+16}, and disseminated through the online 
FRB catalogue. However, the Bayesian parameter estimation and model selection framework that I apply herein 
reveals some potential inaccuracies in previous results. I do not attempt to model FRBs with irregular temporal 
profiles \citep{cpk+16}. I estimate the fluences of the Parkes FRBs by analysing the sky-response model \citep{rsb+16} 
of the thirteen-beam multibeam receiver \citep[MBR;][]{swb+96} used to detect these events. The fluences of 
FRBs detected in individual beams of the MBR may be bounded by making use of their non-detections in other beams. 
Further, for the multiple-beam FRB\,010724 \citep{lbm+07}, I better constrain its fluence by applying the technique developed by 
\citet{rsb+16} for the multiple-beam FRB\,150807. 

Various attempts have been made to infer characteristics of the FRB population by inspecting the distributions of 
FRB properties. Efforts have focused in particular on FRB fluences, DMs and scattering timescales. 
\citet{cfb+16} and \citet{lhz+16} have attempted to use quoted fluences or fluence lower-limits to derive the cumulative fluence 
distribution (the `logN-logF') for FRBs, and thus test for whether the FRB population is consistent with Galactic, nearby extragalactic, 
or cosmological origins. However, \citet{vrh+16} concluded that the FRB logN-logF is best estimated by analysing the numbers of 
single- and multiple-beam detections at Parkes, and by comparing detection rates at different telescopes \citep[see also][]{cpo+16}. 
Additionally, \citet{k16}, \citet{vrh+16} and \citet{cws+16} show that the distribution of FRB DMs is difficult to draw conclusive 
inferences from. However, \citet{k16} and \citet{cws+16} find that the combination of DM and scattering measurements may provide 
an interesting probe of the characteristic host environments of FRBs. Here, I use updated scattering measurements to revisit the 
analysis of Cordes et al. (\S2.4), and find moderate evidence for a relation between FRB DMs and scattering timescales. 

Finally, I discuss the possibility of multiple classes of FRB (\S4). I focus on the comparison 
between the Parkes and Arecibo FRB surveys, which have been undertaken with similar instrumentation, at similar 
frequencies, but with sensitivities and sky-coverages that differ by more than an order of magnitude. It appears possible that 
Parkes could already have detected up to three repeating FRBs like FRB\,121102. Using my analysis of the 
Parkes FRB sample, I speculate on which FRBs may be expected to repeat. I also consider whether all Parkes 
FRBs could be emitted by objects like the source of FRB\,121102, and find that, if so, the population must present 
an incredible diversity of properties. I conclude in \S5.

\section{Modelling of FRB data}

In the sub-sections below, I first describe the FRB data that I model, including the pre-processing steps that I perform (\S2.1). 
I then outline the different models that I attempt to fit to the data, as well as the Monte Carlo Markov Chain 
(MCMC) exploration of the model likelihoods  and the Bayes Information Criterion used to perform model 
selection (\S2.2). The results of the modelling in this section are presented in Table~\ref{table:1}, and discussed in detail in \S2.3. I 
outline the implications of my results for a possible relation between DM and scattering strength in \S2.4. 

\subsection{Description of data and pre-processing}

\begin{table}
\centering
\caption{Properties of FRB detection instruments.}
\begin{tabular}{ccc}
\hline
& AFB & BPSR \\
\hline
Centre frequency (MHz) & 1372.5 & 1382.0 \\
Filterbank & Analogue & Digital \\
& & (4-tap polyphase) \\
Channel bandwidth (MHz) & 3.0 & 0.390625 \\
Number of channels & 96 & 1024 \\
Integration time ($\mu$s) & 125 & 64 \\
Bits per sample & 1 & 2 \\
$\sigma_{i}$ (Jy\,ms) & 0.14 & 0.11 \\
\hline 
\end{tabular}
\label{table:0}
\end{table}

Each Parkes FRB was detected in `filterbank' data recorded for each beam of the 13-beam 21\,cm Multibeam receiver (the MBR). 
Filterbank data are total-power measurements (Stokes I) in numerous spectral channels, 
integrated over sub-millisecond timescales. FRBs 010125 \citep{bb14}, 010621 \citep{ksk+12} and 010724 
\citep{lbm+07}, were detected in data  taken with the Parkes Analogue Filterbank \citep[AFB;][]{mlc+01}, which performed the 
channelisation and integration steps prior to (one-bit) digitisation. The remaining FRBs were detected 
with the Berkeley-Parkes-Swinburne Recorder (BPSR) digital-spectrometer system \citep{kjv+10}, with either `iBOB' or `ROACH' digital signal 
processing cards developed by the Center for Astronomy Signal Processing and Electronics Research (CASPER), 
with identical firmware implementations.\footnote{The open-source CASPER hardware designs and firmware are described online at 
https://casper.berkeley.edu, and reviews of current CASPER developments are detailed by \citet{haa+16}.}  Details of the AFB and BPSR instruments are given in 
Table~\ref{table:0}. 

For all Parkes FRBs besides 131104 \citep{rsj15} and 150807 \citep{rsb+16}, the raw filterbank data are made 
available through the FRB Catalogue \citep{pbj+16}. For FRB\,131104, I made use of raw filterbank data 
that I have direct access to, which I make publicly available with this publication. For FRB\,150807, I make use of 
raw filterbank data made publicly available by \citet{rsb+16}. The DSPSR package \citep{vb11} was used to dedisperse the 
filterbank data and extract 2-second duration PSRCHIVE-format \citep{hvm04} data files at the native time and 
frequency resolutions. Dedispersion was done according to the published DMs for each burst, which I term DM$_{\rm init}$. For the 
BPSR data, channels 0 to 160 (1519.5--1582\,MHz) were excluded in the analysis, because these frequencies were attenuated in the 
analogue signal chain to exclude radio-frequency interference (RFI). 
No further RFI excision was done. Persistent, narrow-band RFI did not significantly affect the analysis both because of the 
level-setting procedures of the AFB and BPSR instruments, and because I also subtracted a mean off-pulse baseline 
level from each channel in the 2-second datasets. Additionally, I searched for significantly time-variable or exceedingly 
strong  (comparable to the system temperature) narrow-band RFI by inspecting the total-power variances of each 
channel in the data, but found none at the $3\sigma$ level. I also found no significant bursts of RFI (narrow- or 
broad-band) coincident with any of the FRBs. 

To accelerate further analysis, the dedispersed data were averaged to four channels within the respective AFB and BPSR 
bands. In each channel, the data were further normalised by the off-pulse standard deviations at a fiducial integration time of 1\,ms. 

FRB\,010724 was detected in four beams of the MBR, and saturated the 1-bit digitiser of the AFB in the 
primary detection beam \citep[beam 6;][]{lbm+07}. I therefore analysed a dataset formed from the sum of the three other detection 
beams (beams 7, 12 and 13).

\subsection{Multi-frequency burst profile modelling}

The pulse-modelling technique employed here closely follows that employed by \citet{rsj15} for FRB\,131104. I used 
a Bayesian technique to find model parameter values that best fit the data, as well as their confidence intervals. 
This technique fully accounts for covariances between model parameters, and allows for accurate parameter confidence intervals 
in the case of non-Gaussian posterior distributions to be presented. I used the {\it emcee} MCMC 
software package \citep{fhl+13} to explore the full likelihood spaces of the multi-frequency models given the data. Following a burn-in stage, 
the joint posterior density of all parameters was estimated with 48000 samples. 

To select 
between models with varying numbers of free parameters, the Bayes Information Criterion (BIC) was calculated for each analysis. The BIC is given by $-2\ln \hat{L}+k[\ln n -\ln(2\pi)]$, where $\hat{L}$ is the likelihood estimate for a model with 
fully specified parameters, $k$ is the number of model parameters, and $n$ is the number of measurements being fit to. 
In accordance with common practise, I selected the model with the lowest BIC, unless there was a model with fewer free parameters with a BIC within three units of the lowest BIC, in which case the model with the fewer free 
parameters was selected. 

The general statistical model that I adopt for the data is outlined in the Appendix. Four specific models were considered, as described below. 

{\bf Model~0.} This model represents a pulse that is temporally unresolved by the instrument, such that the measured 
shape at each frequency is set by the mean intra-channel dispersion smearing of a delta-function impulse. The contribution to the 
measured pulse width from the impulse response of the instrument, quantified approximately as the inverse of the channel 
bandwidth \citep{cm03}, is negligible, and I hence do not include it in the model.  Given a channel 
response function $\tilde{g}(\nu)$ (e.g., Equation A6), where $\nu$ is the radio frequency, the dispersion-smeared temporal profile of a delta-function 
impulse is given by writing $\nu$ in terms of the corresponding dispersion delay (Equation~\ref{eqn:2} below). The 
AFB and BPSR channel responses are both well modelled by Gaussian functions. The model pulse profile in a channel with frequency 
$\nu_{i}$ is given by
\begin{equation}
\label{eqn:1}
S_{i}(t) = \frac{c_{i}}{\sqrt{2\pi\sigma_{i,\,{\rm DM}}^{2}}}\exp\left[\frac{-(t-t_{0}-t_{i,\,{\rm DM}})^{2}}{\sigma_{i,\,{\rm DM}}^{2}}\right],
\end{equation}
where $t_{0}$ is a reference time at the highest frequency, 
\begin{equation}
\label{eqn:2}
t_{i,\,{\rm DM}} = ({\rm 4.15\,ms}){\rm DM_{\rm err}}[(\nu_{i}/{\rm GHz})^{-\beta} - 1.582^{-\beta}]
\end{equation}
with $\beta=2$, and
\begin{equation}
\sigma_{i,\,{\rm DM}} = (1.622\times10^{-3}\,{\rm ms}){\rm DM}(\nu_{i}/{\rm GHz})^{-\beta-1}.
\end{equation} 
Here, ${\rm DM_{\rm err}}$ is the deviation of the burst DM from that assumed in the initial dedispersion of the filterbank 
data (DM$_{\rm init}$), and DM is ${\rm DM}_{\rm init}+ {\rm DM}_{\rm err}$. The coefficients $c_{i}$ are proportional 
to the burst fluences in each of the four frequency channels (indexed by $i$), not accounting for the uncertain positions of the FRBs within the Parkes response function on the sky.  
Representative constants of proportionality, $\sigma_{i}$, assuming beam-boresight positions are given in Table~\ref{table:0} for the AFB and BPSR systems. I 
assume a frequency-independent system temperature of 28\,K, a gain of 1.45\,Jy\,K$^{-1}$, and digitization-loss factors of 0.798 and 0.936 respectively for the AFB and BPSR 
\citep{kp15}.

The model free parameters are therefore the four $c_{i}$ coefficients, $t_{0}$ and ${\rm DM_{\rm err}}$. Note that the assumption 
of $\beta=2$ implies that I assume cold, sparse plasma dispersion; no significant deviations from $\beta=2$ have 
been detected for any FRB, and when relaxing this assumption I also did not find any significant deviations. Although 
the constraining range on $\beta$ can be used to constrain the size of the dispersing region \citep{mls+15}, 
my work does not improve on existing results, and I hence do not report that part of the analysis. 

{\bf Model~1.} This model is the same as Model~0, with the modification of setting $\sigma_{i,\,{\rm DM}}$ to 
$(\sigma_{i,\,{\rm DM}}^{2}+\zeta^{2})^{1/2}$. Here, the new free parameter 
$\zeta$ is an intrinsic burst width. I assume a Gaussian intrinsic profile, in accordance with common practise in modelling 
the mean pulse profiles of pulsars \citep[e.g.,][]{ymv+11}. The quality of the data also do not permit exploration of more 
complex profiles in most cases.  

{\bf Model~2.} This model extends Model~0 by including the effects of temporal broadening due to scattering, clearly detected in some 
FRBs with high signal-to-noise ratios, such as FRB\,110220 \citep{tsb+13} and FRB\,131104 \citep{rsj15}. I account for 
scattering by convolving the temporal profile in Equation~\ref{eqn:1} with a one-sided exponential function:
\begin{eqnarray}
s_{i}(t) &=& \exp\left[\frac{t-t_{0}}{\tau_{\rm 1\,GHz}\nu_{i,1}^{-\alpha}}\right],\,\,\,t>0 \\
&=& 0,\,\,\,{\rm otherwise}.
\end{eqnarray}
Here, $\nu_{i,1}$ is the frequency of channel $i$ expressed in units of 1\,GHz. This form for the pulse-broadening 
function implicitly assumes that the scattering medium can be well approximated by density inhomogeneities projected 
onto a single thin screen \citep{c70}. However, none of the FRB data have sufficient 
sensitivity to distinguish between this and other subtly different forms \citep[e.g.,][]{w72}. The new free parameters are the 
characteristic broadening timescale at 1\,GHz, $\tau_{\rm 1\,GHz}$, and the index of frequency-dependency, $\alpha$.

In some cases, this model was preferred over all others according to the BIC, but the value of $\alpha$ was poorly 
constrained. In these cases, I assumed a value of $\alpha=4$ to estimate $\tau_{\rm 1\,GHz}$, corresponding to 
the expectation for a normal distribution of plasma-density inhomogeneities. In one other case (FRB\,140514), there was 
insufficient sensitivity to distinguish between this model and Model~1. I assumed Model~1 in this case. 

{\bf Model~3.} This model combines Models 0--2, including the effects of both scattering and an intrinsic 
pulse width. 

When no evidence was found for either an intrinsic burst width or scattering, I set upper limits on their values by 
evaluating the posterior distribution for Model~3 with the scattering frequency-dependency index set at $\alpha=4$.

\subsection{Results}

\begin{figure*}
\centering
\label{fig:1}
\includegraphics[scale=0.35]{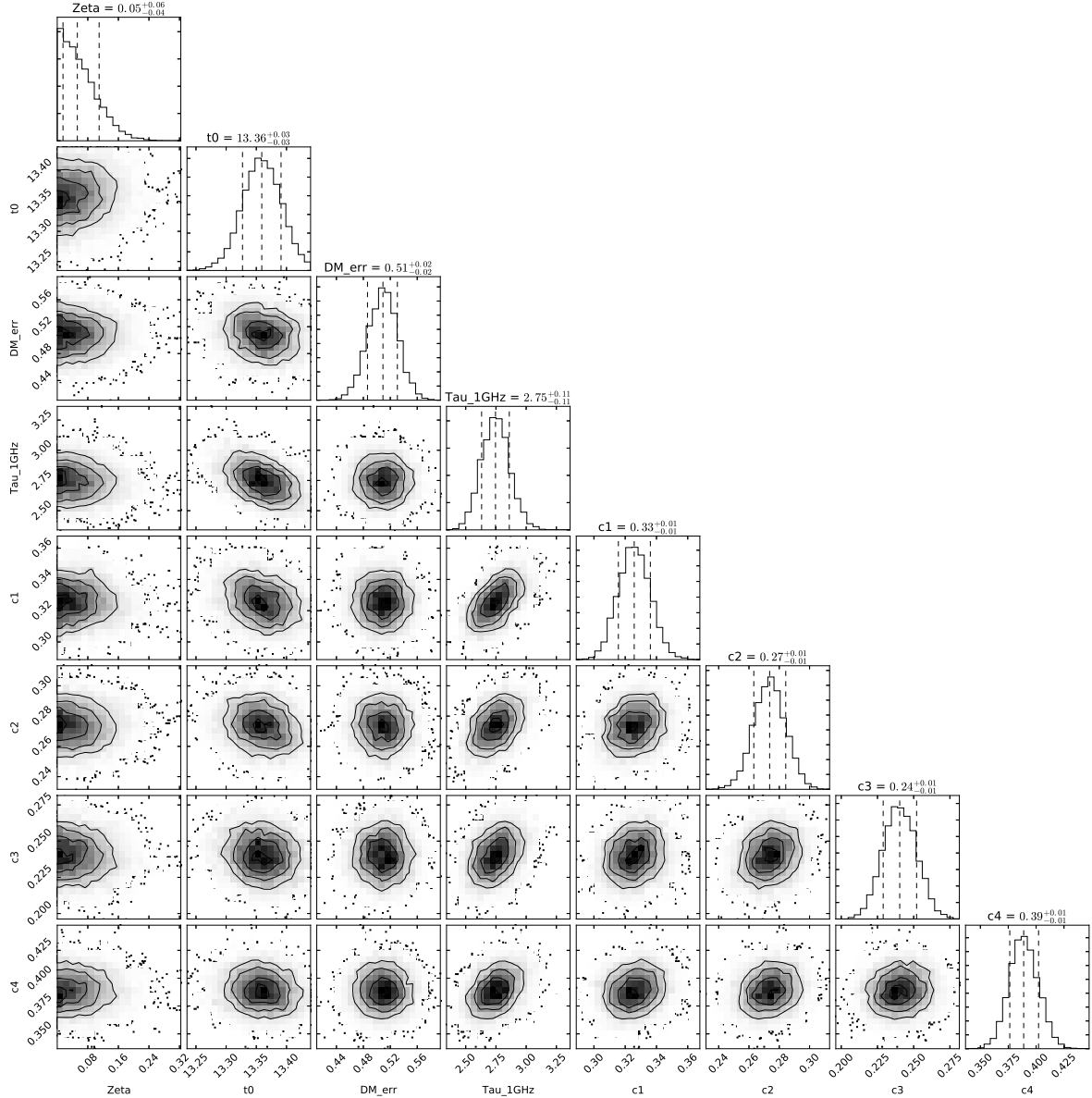}
\caption{Posterior density estimates for a fit of Model~3 to data for FRB\,110220, assuming $\alpha=4$. The parameters 
of the fit were $\zeta$ (Zeta), $t_{0}$ (t0), ${\rm DM}_{\rm err}$ (DM\_err), $\tau_{\rm 1\,GHz}$ (Tau\_1GHz), and $c_{1}$ to 
$c_{4}$ (c1 -- c4). Estimated marginalised posterior densities in each parameter are shown as histograms of samples of the 
posterior, and joint densities between all pairs of parameters are shown by shading and contours. 48000 samples of the posterior 
were obtained.}
\end{figure*}

\begin{table*}
\centering
\caption{FRB properties. Errors in the last significant figures are given in parentheses.}
\begin{tabular}{cccccccccc}
\hline
FRB & DM (pc\,cm$^{-3}$) & $\tau_{\rm 1\,GHz}$ (ms) & $\alpha$ & $\zeta$ (ms) & $c_{1}$ ($\sigma_{1}$) & $c_{2}$ ($\sigma_{2}$) & $c_{3}$ ($\sigma_{3}$) & $c_{4}$ ($\sigma_{4}$) & Best model \\
\hline
010125 & 792.3(1) & $<9.6$ & $-$ & $<1.25$ & 38(4) & 54(5) & 56(4) & 76(5) & 0 \\
010621 & 745.9(2) & $<3.9$ & $-$ & $<1.4$ & 36(4) & 31(4) & 42(5) & 58(5) & 0 \\
010724 & 362.7(1) & 25(5) & 6.4(1.7) & 2.4(3) & 45(4) & 53(4) & 78(5) & 75(5) & 3 \\
090625 & 899.14(6) & 5.2(5) & 4(1) & $<0.2$ & 11.5(7) & 8.2(6) & 9.9(7) & 6.6(7) & 2 \\
110220 & 944.83(5) & 11.4(4) & 3.6(5) & $<0.2$ & 63(2) & 53(2) & 48(2) & 76(2) & 2 \\
110626 & 723.3(4) & $<0.57$ & $-$ & $<0.46$ & 4.5(7) & 5.1(6) & 5.2(9) & 4(1) & 0 \\
110703 & 1104.1(5) & 32(1) & $-$ & $<0.71$ & 8(1) & 20(2) & 19(2) & 12(2) & 2 \\
120127 & 554.22(3) &  $<1.53$ & $-$ & $<0.18$ & 1.1(3) & 1.9(3) & 3.2(4) & 4.8(4) & 0 \\
130626 & 952.01(5) & 2.8(4) & $-$ & $<0.52$ & 5.1(5) & 5.3(5) & 6.6(6) & 5.8(7) & 2 \\
130628 & 469.98(1) & $<0.23$ & $-$ & $<0.04$ & 2.4(1) & 2.1(1) & 1.8(1) & 1.5(2) & 0  \\
131104 & 778.5(1) & 15(2) & 4.4(8) & $<0.18$ & 11.9(6) & 12.5(6) & 10.5(6) & 8.0(6) & 2 \\ 
140514 & 563.8(6) & $<6.1$ & $-$ & 1.2(1) & 7(1) & 8(1) & 14(1) & 16(1) & 1 \\
150215 & 1106.8(3) & $<0.47$ & $-$ & 0.7(1) & 6.1(7) & 5.7(7) & 7.0(8) & 5.6(7) & 1 \\
150418 & 775.84(1) & 0.12(1) & $-$ & $<0.05$ & 1.37(6) & 1.44(7) & 1.57(7) & 1.30(8) & 2 \\
150807 & 266.5(1) & $<0.08$ & $-$ & $<0.04$ & 1.1(1) & 3.58(9) & 12.6(1) & 12.1(1) & 0 \\
\hline 
\end{tabular}
\label{table:1}
\end{table*}

I begin by walking the reader through the fitting process for FRB\,110220, which was modelled by \citet{tsb+13} with a Gaussian profile 
convolved with an exponential scatter-broadened profile. By eye, Models 0 and 1 are inconsistent with the data; 
this was confirmed by exceedingly high BICs for these models. A fit of Model~2 to the 
data resulted in a value of $\alpha=3.6\pm0.5$, which is consistent to within the error range with the \citet{tsb+13} value of $4.0\pm0.4$. 
It is difficult to compare my value of $\tau_{\rm 1\,GHz}=11.4\pm0.4$\,ms with the width at 1.3 GHz estimated by \citet{tsb+13} 
($5.6\pm0.1$\,ms), given differences in the fitted models and the measurement uncertainties. Finally, to check whether any intrinsic width is detectable, I conducted a fit of 
Model~3 to the data with $\alpha$ set to a value of $4$. The estimated posterior densities for this model are shown in Fig.~1. 
It is evident both from the shape of the marginalised posterior distribution 
in $\zeta$, and from a comparison of the BICs between Models 2 and 3 (the BIC for Model~3 was three units greater than the BIC for 
Model~2) assuming the best-fitting parameters, that there is no evidence for an intrinsic width besides the DM-smearing timescale. 
This result is also consistent with the findings of \citet{tsb+13}. 

Fig.~1 also serves to illustrate the levels of covariances that I typically found between model parameters. These are 
 negligible. The scattering timescale, $\tau_{\rm 1\,GHz}$, is most covariant with other parameters, in particular $t_{0}$ and 
$c_{1}$.

\begin{figure*}
\centering
\label{fig:2}
\includegraphics[scale=0.85,angle=-90]{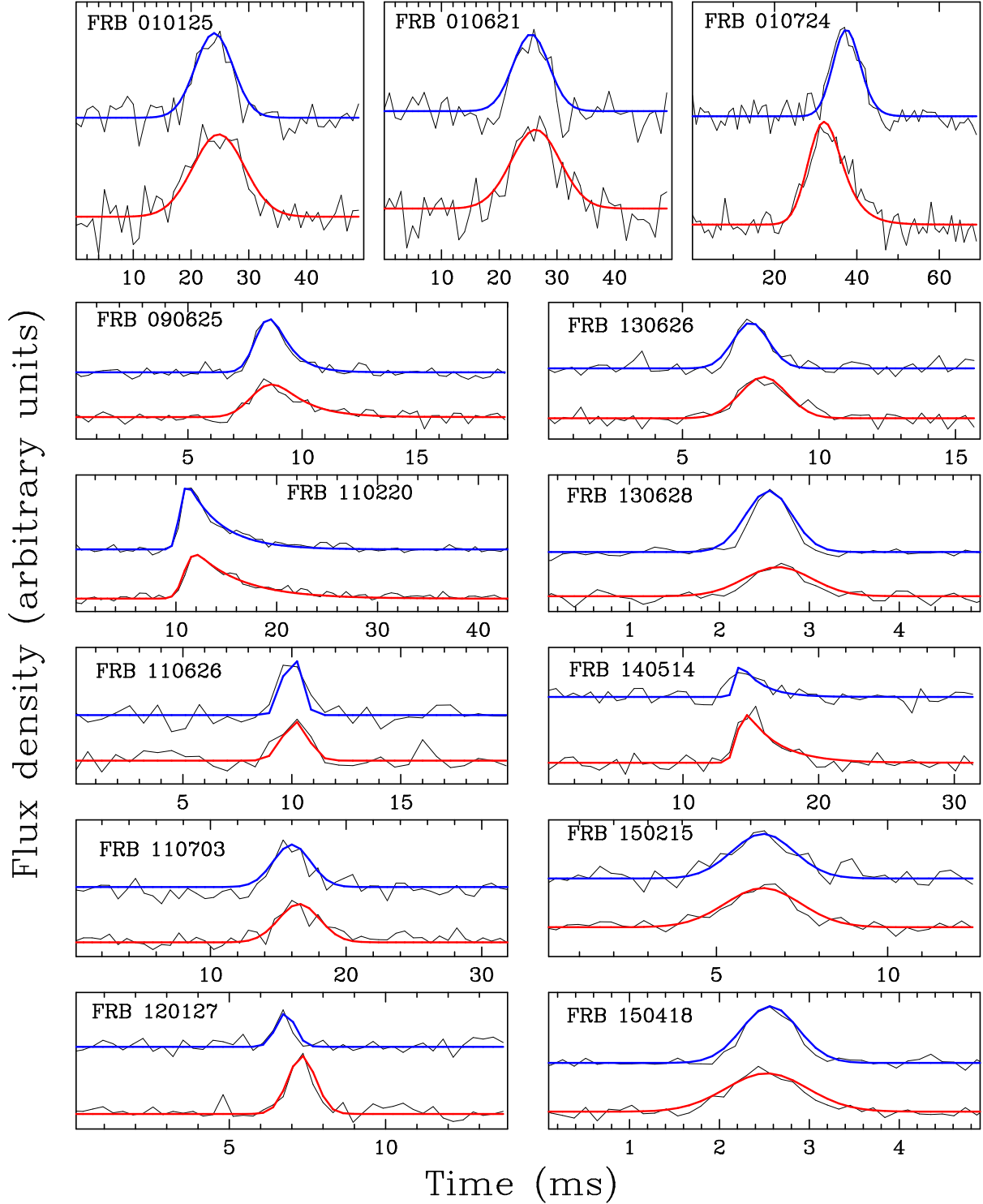}
\caption{Data (thin black lines) and model fits (thick blue and red lines) for a selection of FRBs. In each panel, the top and bottom curves 
(corresponding to the blue and red dashed lines respectively) are the mean temporal profiles of the FRBs in the upper- and lower-frequency 
halves of the observing bands, respectively. The relative flux-density scales between the temporal profiles in the two bands are normalised to 
the respective noise levels; the uniform frequency-response of the Parkes multibeam system implies that the temporal profiles in 
the two bands are on approximately the same absolute amplitude scale. Details of the exact bandwidths are given in the text, and 
details of the model fits are presented in Table~\ref{table:1}.}
\end{figure*}


I show fits to data on thirteen of the Parkes FRB sample in Fig.~2; details of the specific models and 
best-fit parameters are given in Table~\ref{table:1}. I show temporal profiles averaged over the upper and lower halves 
of the respective observing bands (\S2.1), after rejection of channels 0--160 in the case of BPSR.  I dedispersed the data using the DMs from the original analyses of the FRBs (DM$_{\rm init}$); 
in some cases, significantly different DMs were derived (e.g., FRB\,010724; Fig.~2 top-right panel). I do not show the 
results for FRBs 131104 and 150807, because they have been previously fit using my technique \citep{rsj15,rsb+16}. 
Note that in Table~\ref{table:1}, the fluence coefficients $c_{i}$ are scaled to be in units of $\sigma_{i}$-ms, where $\sigma_{i}$ is the 
noise-floor standard deviation in channel $i$ in one millisecond (see Table~\ref{table:0}). 

The final two Parkes FRBs that are excluded from Fig.~2, 121002 and 130729, could not be modelled using any of Models 0--3. 
These FRBs are also excluded from Table~\ref{table:1}. This is because they both exhibit two temporal components. I show the 
dedispersed dynamic spectra of these FRBs in Fig.~3. Interesting spectral structure is also present in FRB\,130729, which appears 
concentrated in the lower part of the observing band. 

\begin{figure*}
\centering
\label{fig:4}
\includegraphics[scale=0.55]{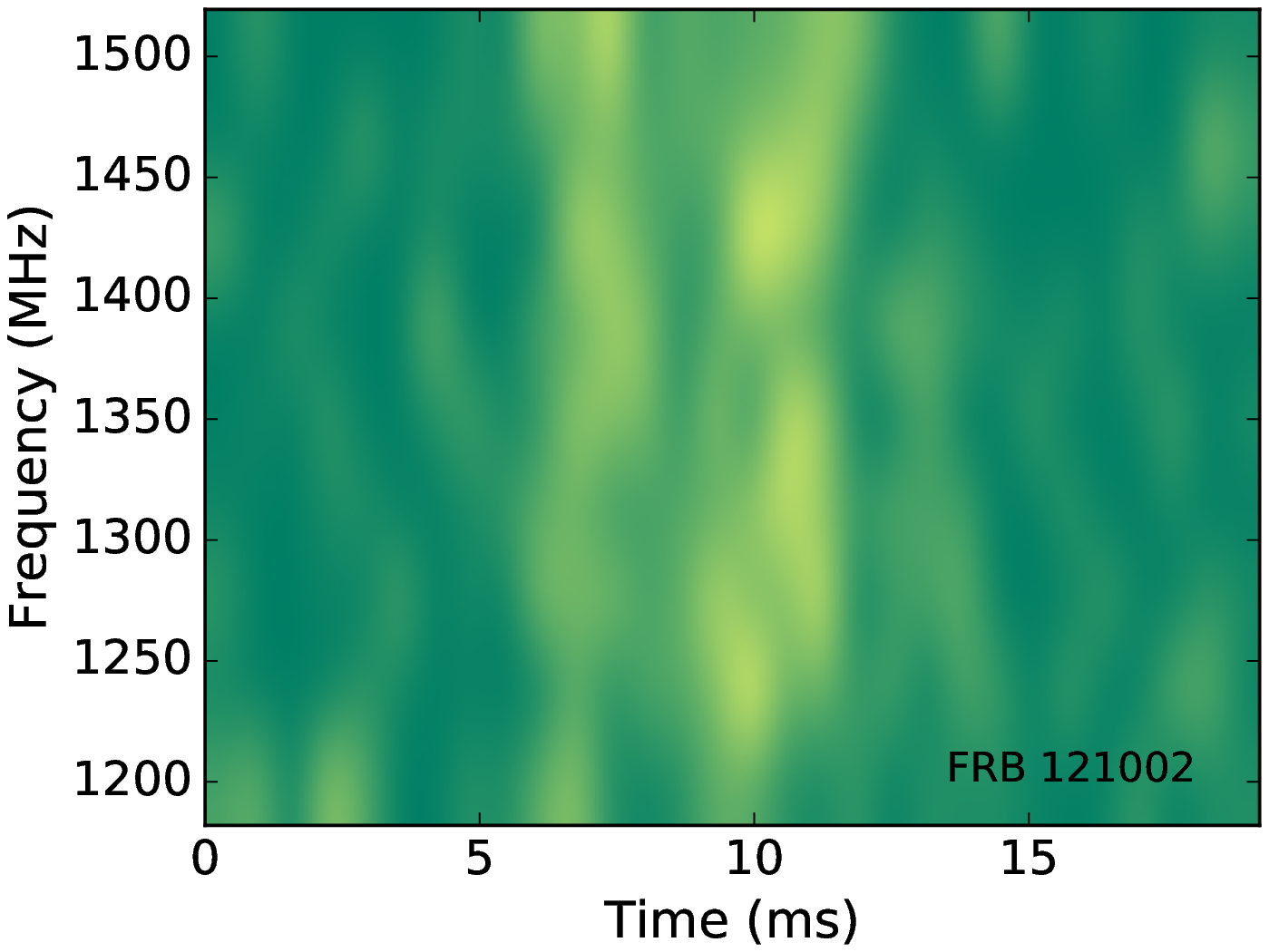}
\includegraphics[scale=0.55]{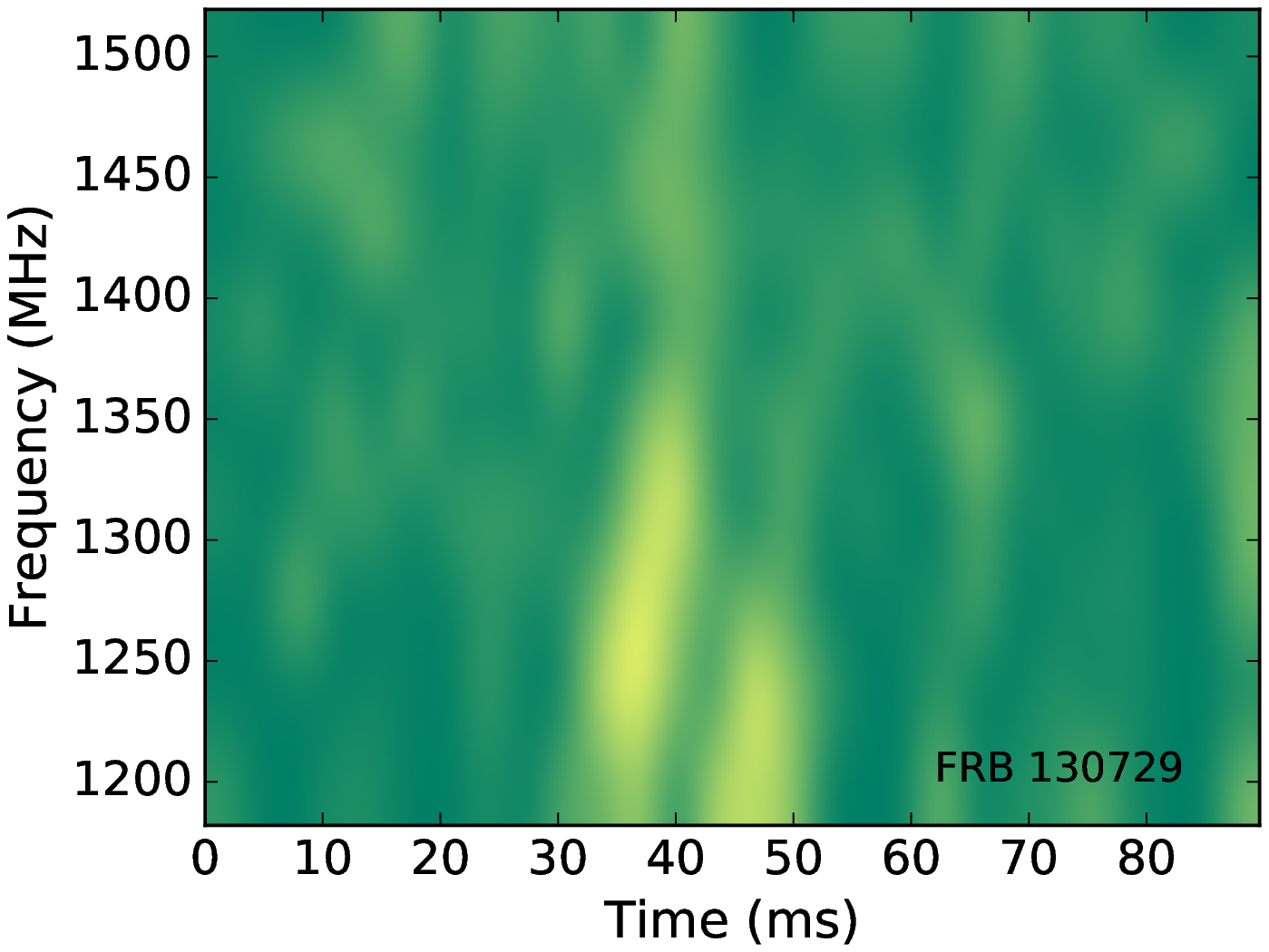}
\caption{Dedispersed dynamic spectra of FRBs 121002 (left) and 130729 (right), for which Models 0--3 were insufficient. In both cases, 
two temporal components are evident. The dynamic spectra have been interpolated using a bicubic spline fit. As with Fig.~2, 
the data in each spectral channel have been normalised to the respective noise levels; the colour-scale therefore represents the signal to noise ratio (S/N) 
in linear units.}
\end{figure*}

\subsubsection{Notes on individual FRBs}

{\bf FRB\,010125}: \citet{bb14} found a width for this FRB of $\zeta\approx 5$\,ms, in excess of the DM smearing timescale, although it 
was unclear from their analysis whether this was intrinsic to the pulse or caused by scattering. By analysing the variation with 
frequency of the pulse width, they claim a detection of scattering with $\alpha=4.2\pm1.2$. This is also 
consistent with $\alpha=3$, which would simply correspond to a DM-smeared pulse, as they did not appear to account for DM 
smearing in their analysis of the frequency-variation of the pulse width. My analysis suggests that this was indeed the case: I find no 
evidence for temporal structure in FRB\,010125 besides DM smearing (Model~0). 

{\bf FRB\,010621}: In agreement with \citet{ksk+12}, the present analysis reveals no evidence for temporal structure in this FRB besides DM smearing. Although the 
Galactic disk DM contribution along this low-Galactic-latitude sightline is expected to be 534\,pc\,cm$^{-3}$, the expected 
scattering timescale is only $\tau_{\rm 1\,GHz}=0.15$\,ms \citep{cl02}; this is well below our upper limit of $\tau_{\rm 1\,GHz}<3.9$\,ms 
(95\% confidence). 

Through an analysis of velocity-resolved H$\alpha$ and H$\beta$ observations of the Galactic interstellar medium (ISM) along the 
burst sightline, \citet{bm14} concluded that previous estimates for the Galactic disk DM contribution were underestimated, 
and that this burst is in fact Galactic (90\% confidence). A potential problem for this hypothesis is my upper limit on the 
scattering timescale. \citet{bm14} predict a scattering timescale of $\approx2.4$\,ms in the observing band, corresponding to 
$\tau_{\rm 1\,GHz}\approx8.5$\,ms, which is excluded by my upper limit. On the other hand, the relation between DM and 
$\tau_{\rm 1\,GHz}$ in the Galaxy has a large intrinsic scatter \citep[0.76\,dex;][]{cws+16}. Nonetheless, for a Galactic sightline with 
the DM of the burst (746\,pc\,cm$^{-3}$), the burst would have to be under-scattered by a factor of $\approx2.5\sigma$. 
This could be because significant amounts of DM are contributed by higher-density gas surrounding the source or hot ISM with 
weak density fluctuations, or that the scattering is dominated by localised clumps rather than the bulk ISM \citep{cws+16}.  

{\bf FRB\,010724}: I find moderate evidence for both an intrinsic width and an exponential scattering ``tail'' in this FRB. The present analysis 
differs from that of \citet{lbm+07} because it uses the sum of data from the three non-saturated beams, rather than data from the 
saturated beam alone. Nonetheless, my estimates of the scattering timescale, $\tau_{\rm 1\,GHz}=25\pm5$\,ms, and index, 
$\alpha=6.4\pm1.7$, are consistent with those of \citet{lbm+07}  ($24.13\pm3$\,ms and $4.8\pm0.4$ respectively). I also revise the 
DM estimate from 375\,pc\,cm$^{-3}$ to $362.7\pm0.1$\,pc\,cm$^{-3}$. 




{\bf FRB\,110703}: Unlike the analysis of \citet{tsb+13}, I find moderate evidence for the presence of a significant scattering tail in this FRB 
($\tau_{\rm 1\,GHz}=32\pm1$\,ms). However, a constrained value for $\alpha$ cannot be determined, and I hence assume $\alpha=4$. 




{\bf FRB\,130729}: Like FRB\,121002, this burst has two temporal components. Unlike FRB\,121002, FRB\,130729 also has a 
discontinuous spectrum (Fig.~3), with most power concentrated in the lower part of the band. It is unclear whether the 
scattering timescale derived by \citet{cpk+16} for this FRB is real, or is attributable to the unusual temporal and spectral structure. 

{\bf FRB\,130628}: Unlike \citet{cpk+16}, I find no evidence for a scattering tail in this FRB, or for any structure beyond 
that described by Model~0. Indeed, my upper limit on the scattering timescale, $\tau_{\rm 1\,GHz}<0.23$\,ms, is well below 
the previous estimate of $\tau_{\rm 1\,GHz}=1.24\pm0.07$\,ms. 


{\bf FRB\,140514}: For this FRB, Models 1 and 2 had equivalent BICs. I hence choose Model~1 for this FRB, and thus do not 
find any evidence for the existence of scattering, unlike \citet{pbb+15}. Along with FRBs 121002 and 130729, this is one of the 
few FRBs to exhibit temporal structure beyond Model~0 with no clear evidence of scattering. It also has a mildly inhomogeneous 
spectrum, as indicated by the $c_{i}$ coefficients in Table~\ref{table:1}. 

{\bf FRB\,150215}: In agreement with \citet{pbk+17}, I find evidence for a larger temporal width than is expected in Model~0, which appears 
to be intrinsic to the burst. 

{\bf FRB\,150418}: In contrast to the analysis of \citet{kjb+16}, I find that this FRB exhibits a weak scattering tail, with a timescale of 
$\tau_{\rm 1\,GHz}=0.12\pm0.01$\,ms. The value of $\alpha$ cannot be constrained. 

{\bf FRB\,150807}: This FRB was modelled using similar techniques by \citet{rsb+16}. No evidence was found for any temporal 
structure beyond Model~0 in either the previous or present analysis. Although I present an upper limit on $\tau_{\rm 1\,GHz}$ 
in Table~\ref{table:1}, in the analysis below I use the value inferred from the frequency-scintillations of $1.6\pm0.8\,\mu$s \citep{rsb+16} at 
1.3\,GHz, corresponding to $\tau_{\rm 1\,GHz}=4.6\pm2.3\,\mu$s. 

\subsection{Astrophysical implications}

\begin{figure}
\centering
\label{fig:7}
\includegraphics[scale=0.55,angle=-90]{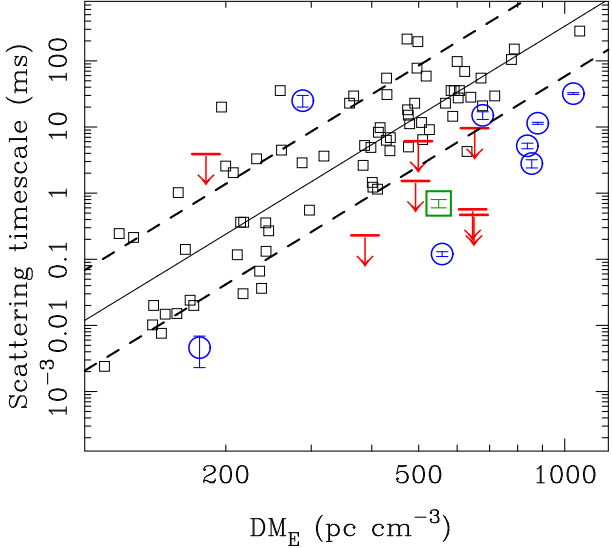}
\caption{Measurements of ${\rm DM}_{E}$ and $\tau_{\rm 1\,GHz}$ for FRBs in Table~\ref{table:1}: cases where 
$\tau_{\rm 1\,GHz}$ is measured are shown as blue circles with error-bars, and upper-limits on $\tau_{\rm 1\,GHz}$ are indicated 
by downward-facing red arrows. I also show measurements for the Green Bank Telescope FRB\,110523 \citep{mls+15} as a large green square. 
The solid and dashed lines indicate the Galactic $\tau_{\rm 1\,GHz}$-DM relation and its intrinsic scatter, 
respectively, derived by \citet{cws+16}. The black squares show pairs of measurements of DM and $\tau_{\rm 1\,GHz}$ for 
Milky-Way pulsars from the ATNF pulsar catalogue \citep{mht+05}. All pulsars with published measurements of $\tau_{\rm 1\,GHz}$ are included here.}
\end{figure}

As foreshadowed in the Introduction, an immediate utility of my quantitative results is to investigate the scattering strengths of FRBs at 
different DMs. I have only marginally adjusted the FRB DMs, and, as shall be shown in the following section, the fluence constraints for 
most FRBs are not tight enough to enable a rigorous analysis of the distribution of FRB fluences. 

A relationship between the scattering timescale, $\tau_{\rm 1\,GHz}$, and DM is firmly established for Milky-Way pulsars over 
three orders of magnitude in DM, and eleven orders of magnitude in $\tau_{\rm 1\,GHz}$. The large intrinsic scatter of 0.76\,dex, and 
the steeper slope of the relation at ${\rm DM}\gtrsim100$, are interpreted as evidence for clumpiness in the ionised ISM 
\citep{cwf+91,cws+16}. Motivated by the strong evidence for scattering in FRB\,110220 presented by \citet{tsb+13}, the possibility of 
a $\tau_{\rm 1\,GHz}$-DM relation existing for FRBs putatively scattered in the intergalactic medium (IGM) was first considered by 
\citet{lkm+13}. However, the possibility of any significant scattering in the IGM was disputed by \citet{mk13} and \citet{lg14}, based on 
their assessments of IGM turbulence. 

The existence of a $\tau_{\rm 1\,GHz}-{\rm DM}_{E}$ relation for FRBs, where ${\rm DM}_{E}$ is the estimated extragalactic 
DM component for a given FRB, would thus imply that FRBs are predominantly scattered in the ionised medium that dominates 
the ${\rm DM}_{E}$ values. For example, if ${\rm DM}_{E}$ is typically dominated by contributions 
from the IGM, it would be possible that FRBs are predominantly scattered in the IGM or in intervening bound systems. 
On the other hand, if ${\rm DM}_{E}$ is typically dominated by host-galaxy contributions, a $\tau_{\rm 1\,GHz}-{\rm DM}_{E}$ relation 
would reflect the typical DM-$\tau_{\rm 1\,GHz}$ relation in FRB host galaxies. The lack of a 
$\tau_{\rm 1\,GHz}-{\rm DM}_{E}$ relation would imply that the medium that dominates the ${\rm DM}_{E}$-values does not 
significantly scatter FRBs. 

\citet{cws+16} compared existing measurements of FRB scattering timescales with a revised $\tau_{\rm 1\,GHz}$-DM relation for 
Milky-Way pulsars. No evidence was found for a $\tau_{\rm 1\,GHz}-{\rm DM}_{E}$ relation for FRBs. However, 
it was shown that FRBs are typically under-scattered in comparison to their values of ${\rm DM}_{E}$, relative to Milky-Way 
pulsar scattering timescales at congruent values of DM. This was interpreted either as an indication that $50-75\%$ of ${\rm DM}_{E}$ 
is typically contributed by the IGM, or that FRB host galaxies have ISMs that are typically less turbulent than the ISM of the Milky Way. 
The possibility of FRBs being predominantly scattered in the IGM was thought less likely owing to the large levels of scattering 
present relative to expectations for the IGM \citep[e.g.,][]{mk13}, and the lack of a $\tau_{\rm 1\,GHz}-{\rm DM}_{E}$ relation. 

The discovery of FRB\,150807 \citep{rsb+16}, however, significantly extended the range of FRB scattering timescales. Although not 
detectably temporally broadened due to scattering, this FRB exhibited frequency-scintillations that indicated a scattering strength much 
greater than that expected from its Milky-Way sightline (Shannon et al., in preparation). The combined measurements of low 
scattering and low Faraday rotation-measure 
indicated that this FRB was most likely not scattered in ISM with turbulence and magnetisation like that of the Milky Way. However, 
constraints on the distance to the source of FRB\,150807 suggested that a significant portion of its ${\rm DM}_{E}$ originated in the 
IGM. 

Using my revised measurements of DM and $\tau_{\rm 1\,GHz}$ for the Parkes FRB sample (Table~\ref{table:1}), I plot 
$\tau_{\rm 1\,GHz}$ against ${\rm DM}_{E}$ in Fig.~4. I also show measurements for FRB\,110523 \citep{mls+15} discovered at the 
Green Bank Telescope in the 700-900\,MHz band. 
For each FRB, I estimate ${\rm DM}_{E}$ by subtracting the maximum 
Galactic-disk DM predicted by the NE2001 DM model \citep{cl02} along the FRB sightline, and by further subtracting a contribution of 
30\,pc\,cm$^{-3}$ corresponding to the Milky Way ionised-gas halo and the Local Group \citep[e.g.,][]{gmk+12,dgb+15}. In Fig.~4, I also plot the Milky Way 
$\tau_{\rm 1\,GHz}$-DM relation derived most recently by \citet{cws+16}, and pairs of measurements of $\tau_{\rm 1\,GHz}$ and DM 
for Milky-Way pulsars from the ATNF pulsar catalogue \citep{mht+05}. 

First, Fig.~4 supports the finding of \citet{cws+16} that FRBs are under-scattered with respect to their values of 
${\rm DM}_{E}$. This is despite the differences in the actual measurements, and in the compositions of the samples, between the 
two analyses. Relative to the Cordes et al. analysis, I discard FRBs 121002 and 130729 due to their complex temporal and 
spectral structures, which may have biased previous scattering measurements, but include the new FRB\,150807 and its 
value of $\tau_{\rm 1\,GHz}$ based on the frequency-scintillations. The exclusion of FRBs 121002 and 
130729 may bias inferences from Fig.~4, because our ability to discern the complex temporal structure relies on them not being 
strongly scattered. However, any upper limits that could be placed on their scattering timescales would correspond approximately to the 
narrowest features in the burst profiles (i.e., a few milliseconds at $\sim1.3$\,GHz; see Fig.~3). These in turn would be approximately 
consistent with existing measurements in Fig.~4. 
 
Fig.~4 provides tentative indications of a relation between $\tau_{\rm 1\,GHz}$ and ${\rm DM}_{E}$  for FRBs similar to that for 
Milky-Way pulsars. If FRB\,010724, which has the largest $\tau_{\rm 1\,GHz}/{\rm DM}_{E}$ ratio among the FRB sample, is excluded, 
the $\tau_{\rm 1\,GHz}-{\rm DM}_{E}$ relation appears somewhat stronger. Using the BIC, I quantify the evidence for a 
$\tau_{\rm 1\,GHz}-{\rm DM}_{E}$ relation by comparing linear models for the ${\rm DM}_{E}$ and $\tau_{\rm 1\,GHz}$ measurements, 
with, and without, a dependency of $\tau_{\rm 1\,GHz}$ on ${\rm DM}_{E}$. Consider the log-likelihood function 
\begin{equation}
L = \sum_{i} \left[-\log(\epsilon_{i}^{2}+\epsilon^{2})-\frac{(\log_{10}\tau_{\rm 1\,GHz,i}-M_{i})^{2}}{2(\epsilon_{M,i}^{2}+\epsilon^{2})}\right]
\end{equation}
where $M_{i} = m\log_{10}{\rm DM}_{E,i}+b$ is a log-linear model for the $\tau_{\rm 1\,GHz}-{\rm DM}_{E}$ relation with parameters $m$ and $b$ 
and intrinsic scatter $\epsilon$, and ${\rm DM}_{E,i}$ and $\tau_{\rm 1\,GHz,i}$ are FRB measurements indexed by $i$ with error $\epsilon_{i}$. 
I consider the difference in BIC between the maximum of this likelihood function in the parameters $m$, $b$ and $\epsilon$, and 
the maximum of the 
likelihood function with a fixed $m=0$ (and hence one less parameter). With the sample of eight Parkes FRBs with measurements of 
$\tau_{\rm 1\,GHz}$, there is no significant difference in the BICs between the two models. However, with FRB\,010724 excluded, 
the difference in BICs is 8, which I consider moderately significant. The inclusion of the upper-limits on scattering timescales does not 
significantly alter these results. For the seven-FRB sample (excluding FRB\,010724), I find $m=7\pm2$ and $b=-19\pm5$; I 
emphasise that these results are likely to change significantly as more scattered FRBs are discovered. 

A $\tau_{\rm 1\,GHz}-{\rm DM}_{E}$ relation for FRBs is would not be particularly surprising, because it would simply imply that significant 
portions of FRB DMs are contributed by a class of medium that has a scattering strength which scales with its column-density. 
It is well-established that 
the Milky Way ISM is one such class of medium. It is generally thought to be unlikely that FRBs are predominantly scattered in 
the Milky Way itself, because they would lie along sightlines of intolerably large $\tau_{\rm 1\,GHz}$ for the Milky-Way DM contributions. 
If FRBs were, however, scattered in host galaxies like the Milky Way, approx. 75\% of the 
typical FRB DM must be contributed by an IGM that has a weak potential for scattering. This is difficult to reconcile with the results on 
FRB\,150807 (Ravi et al. 2016). Note further that in this case a fair comparison between 
the Milky-Way DM-$\tau_{\rm 1\,GHz}$ relation and FRB measurements would require the values of $\tau_{\rm 1\,GHz}$ for Milky 
Way sightlines to be scaled up by a factor of three \citep{cws+16}, to account for the difference in scattering geometry between the 
Milky Way 
(presumably a homogeneous scattering medium along the line of sight), and FRBs scattered in host galaxies (scattering medium 
concentrated around the FRBs). Alternatively, FRBs may instead be scattered in the IGM, or in intervening bound 
systems, and experience negligible host-galaxy scattering. An attempt to ascertain the necessary properties of scattering regions in the 
IGM and intervening systems is beyond the scope of this work. The different scenarios for FRB scattering will be tested when multiple scattered FRBs are 
localised to individual host galaxies, and their distances thus measured, such that the host and IGM contributions to the DM may 
be separately estimated. In any scenario for the dominant contributor of FRB DMs, a $\tau_{\rm 1\,GHz}-{\rm DM}_{E}$ implies that more 
(cosmologically) distanct FRBs will be more difficult to detect. 

\section{FRB flux densities}

Here, I quantify the constraints that may be placed on FRB fluences based on an analysis of the Parkes MBR sky-response. In \S3.1, 
I consider what constraints may be placed on the flux densities of FRBs detected in individual beams of the MBR. Then, in \S3.2, 
I constrain the location of the multiple-beam FRB\,010724 in the Parkes focal plane using a technique similar to that 
applied by \citet{rsb+16} to the dual-beam FRB\,150807. Third, in \S3.3, I combine these analyses with the fluence estimates 
presented in Table~\ref{table:1}, and compare the resulting FRB fluence constraints with various specifications for the FRB 
fluence distribution (the logN-logF). 

\subsection{Single-beam FRBs}

\begin{figure*}
\centering
\label{fig:5}
\includegraphics[scale=0.65]{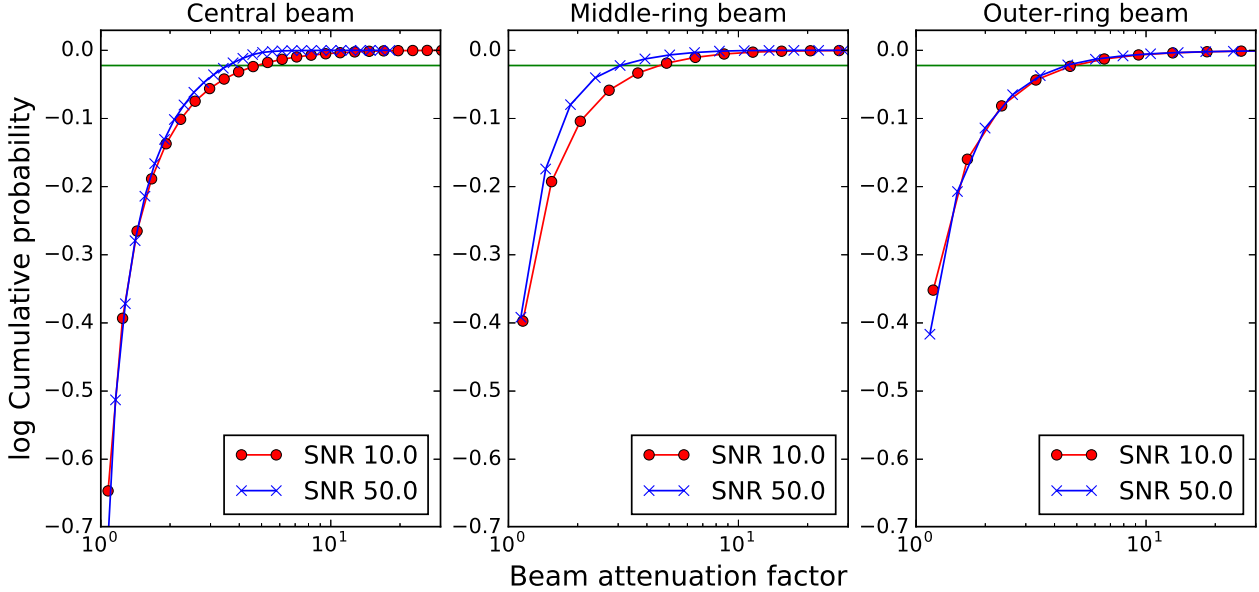}
\caption{The cumulative probabilities, $P(>\Theta)$, for FRBs being detected above different beam attenuation factors, $\Theta$ 
(see text for details). 
From left to right, I show results for the central, an inner-ring, and an outer-ring beam of the Parkes MBR. I 
consider FRBs detected with S/N of 10 (red line with filled circles) and 50 (blue line with crosses). The horizontal green lines 
indicate the 95th percentiles of the distributions.} 
\end{figure*}

The exact locations of single-beam FRBs within the sky-response functions of the beams, $\Theta(\theta,\phi)$, 
are unknown. Here, $\Theta(\theta,\phi)$ is the attenuation of the FRB due to an off-axis position at polar coordinates, $(\theta,\phi)$, 
in the focal plane; for an on-axis feed, $\Theta(\theta,\phi)$ may be well-approximated by the inverse of an Airy function. That is, the 
beam attenuation factor is the inverse of the standard beam gain pattern. By 
adopting models for $\Theta(\theta,\phi)$ for the Parkes receiver, and for the distribution of FRB fluences, it is possible to evaluate 
the probable beam-attenuations of FRBs detected in different beams.

I used the model for $\Theta(\theta,\phi)$ for the Parkes MBR presented by \citet{rsb+16}. This analytic model was found to be consistent 
with measurements at the $-20$\,dB response level. For each beam, I used the model to evaluate $\Theta(\theta,\phi)$ on a grid of $1000\times1000$ 
points spanning $3\times3$\,deg in $\theta$ and $\phi$. I averaged the model in frequency across the BPSR band ($1182-1519.5$\,MHz); 
I did not find the results in this section to vary significantly when instead using the AFB band. Then, for the central, an inner-ring, and an outer-ring 
beam, I derived the histograms of pixel-values of $\Theta(\theta,\phi)$ where the FRB would not be detected with  ${\rm S/N}>3$ in any 
other beam; I considered S/Ns of 10 and 50 in the primary detection beams. These histograms provided initial estimates of the 
probability density functions of $\Theta$ for FRBs detected in individual beams of the MBR.

However, FRBs are not equally likely to be detected at different fluences or flux-densities\footnote{In this context, fluence and flux density 
can be used interchangeably.}: the specific distribution in these parameters is the logN-logF function. For a fluence $F$, the 
number of FRBs expected at fluences $>F$ is typically modelled as a power-law: $N(>F)\propto F^{-\beta}$, for some power-law index 
$\beta$ \citep[e.g.,][]{vrh+16}.  For a uniform distribution of FRBs in Euclidean space, $\beta=1.5$. However, based on the unexpected 
detection of multiple-beam FRBs at Parkes (FRBs 010724 and 150807), \citet{vrh+16} showed that the Parkes FRB sample is consistent 
with $0.5<\beta<0.9$ (90\% confidence). I therefore adopted $\beta=0.7$, and scaled the $\Theta$-histogram counts accordingly. Finally, I 
used the histograms to derive the probabilities of detecting FRBs above given values of $\Theta$ (Fig.~5). 

From Fig.~5, it is apparent that single-beam FRBs in any beam are most likely to be detected with $\Theta<5$ ($>95\%$ confidence 
for ${\rm S/N}\lesssim50$). They are expected to be detected with $\Theta<2$ with $\approx80\%$ confidence, suggesting that the 
conventionally quoted FRB localisation accuracy of the FWHM of the primary beam \citep[e.g.,][]{pbb+15,kjb+16} is only moderately efficacious. 
If the slope, $\beta$, of the FRB logN-logF function were larger(smaller), the maximum likely value of $\Theta$ would decrease(increase). 
For higher-significance FRBs, the maximum likely value of $\Theta$ would also decrease, although only marginally so for FRBs detected 
in outer-ring beams.

\subsection{FRB\,010724 (the Lorimer burst)}

\begin{figure*}
\centering
\label{fig:6}
\includegraphics[scale=0.57]{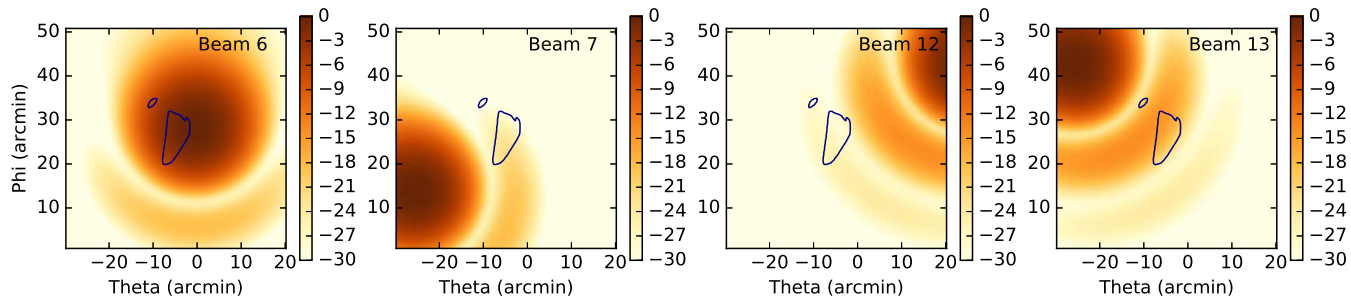}
\caption{The 99\% confidence containment region of FRB\,010724 (blue contours) shown relative to the sky-response functions of beams 6, 7, 12 and 13 
of the Parkes MBR. The colour-bars indicate the beam attenuation factors, $\Theta(\theta,\phi)$, at different positions in the 
Parkes focal plane, in units of decibels.}
\end{figure*}

As discussed above, FRB\,010724 was detected in four beams of the Parkes MBR: the inner-ring beams 6 and 7, and the outer-ring beams 
12 and 13. Following the technique of \citet{rsb+16}, I use the relative S/Ns of the burst in the different beams, and a model for the 
individual sky-response functions of the beams, to constrain the position of the burst in the Parkes focal plane. This in turn provides a 
constraint on the flux-density of the burst. I defer an analysis of the localisation region on the sky of FRB\,010724 to a future paper.

FRB\,010724 saturated the one-bit AFB digitiser for beam 6, making it impossible to accurately measure the S/N in this beam. 
Accurate measurements are however possible for beams 
7, 12 and 13. After averaging the data for these beams over the AFB band, I smoothed each time-series with a top-hat function of width 6\,ms, 
and estimated the peak S/N in each beam. I ensured that the peak S/N in each beam occurred at the same time. For beams 7, 12 and 13, the S/Ns were 
13.9, 5.5 and 22.1 respectively. Unlike the analysis of FRB\,150807 by \citet{rsb+16}, I did not attempt to use measurements at 
multiple frequencies because of the relatively weak detections in these three beams. 

I evaluated models for the sky-response functions of each beam, averaged over the AFB band, using the publicly-available codes presented by 
\citet{rsb+16}. The models were evaluated, as above, on a grid of $1000\times1000$ points spanning $3\times3$\,deg in the Parkes focal plane. 
I then made Monte Carlo realisations of the S/N in each beam, based on the estimated S/Ns, to calculate a containment region. For each realisation, 
I found points in the focal plane where the ratios of S/Ns between beams 7, 12 and 13 were within a factor of four of the simulated measurements. 
I accounted for the difference in telescope gain between inner- and outer-ring beams. 
I rejected points where the burst would have been detected with ${\rm S/N}\geq3$ in any of beams 1, 2, 3, 4, 5, 8, 9, 10 and 11, and also rejected points 
where the burst would have been detected with a lower S/N in beam 6 than in any other beam. Finally, I averaged the results over 1000 realisations. I note that 
this analysis places no prior on the logN-logF function, unlike the analysis in \S3.1. 
 
In Fig.~6, I show the resulting 99\% confidence containment region for FRB\,010724 in the Parkes focal plane. The localisation is to approx. 50 
arcmin$^{2}$. This is substantially worse than the 9-arcmin$^{2}$ localisation of FRB\,150807. This is most likely because I do not use measurements at 
different frequencies for FRB\,010724, whereas measurements in four sub-bands were used to localise FRB\,150807. 

The results suggest that the S/Ns in beams 7, 12 and 13 should be scaled upwards by factors of $310\pm180$, $510\pm140$ and $80\pm40$, respectively. 
Based on the S/N measurements, the fluence measurements for FRB\,010724 in Table~\ref{table:1} should thus be scaled upwards by factor of $200\pm100$. 
Assuming gains of 1.45\,Jy\,K$^{-1}$ and 1.72\,Jy\,K$^{-1}$ for the inner- and outer-ring beams, a common system temperature of 
28\,K \citep{mlc+01}, and a one-bit digitisation loss factor of 1.25 \citep{kp15}, the mean fluence of FRB\,010724 within the AFB band is 
$(800\pm400)$\,Jy\,ms.

\subsection{Astrophysical implications}

\begin{figure}
\centering
\label{fig:8}
\includegraphics[angle=-90,scale=0.49]{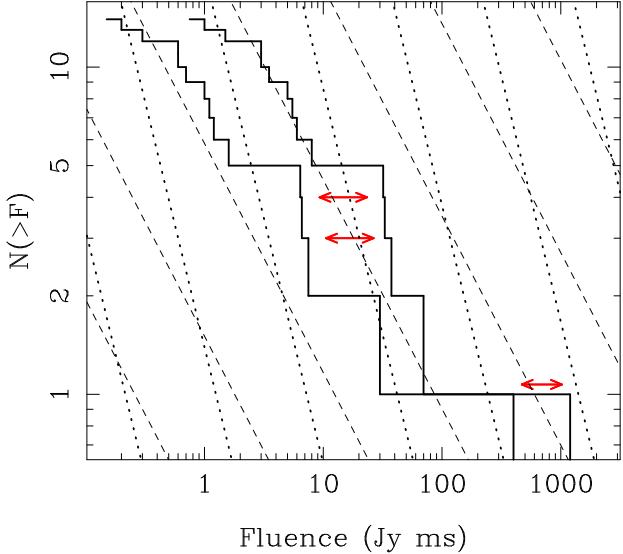}
\caption{Empirical logN-logF distribution for the Parkes FRB sample that I model herein. The lower-fluence binned trace indicates the 
minimum fluence values for each FRB, and the higher-fluence binned trace indicates the maximum fluence values for each FRB. 
The minimum fluences are derived assuming boresight positions in the detection beams for most FRBs, and the lower bounds on the 
$1\sigma$ fluence ranges for FRBs 010724 and 150807. The maximum fluences are the 95\% confidence upper limits on the fluences 
for most FRBs derived using the results in Fig.~5, and the upper bounds on the $1\sigma$ fluence ranges for FRBs 010724 and 
150807. The dashed lines indicate logN-logF functions proportional to $F^{-0.7}$ \citep{vrh+16}, and the dotted lines 
indicate logN-logF functions proportional to $F^{-3/2}$. Although the fluence upper-bounds 
derived using Fig.~5 assume a logN-logF function proportional to $F^{-0.7}$, the higher-fluence binned trace would simply be 
shifted lower or higher in fluence for steeper or flatter logN-logF functions, respectively. The horizontal red arrows indicate the FRBs 
detected with the AFB.}
\end{figure}

The best utility of my revised fluence constraints for the Parkes FRB sample is to consider the implications for the logN-logF. 
I use the fluence estimates in Table~\ref{table:1} for all FRBs besides 010724  and 150807, averaged over the frequency bands, and the 
sensitivity parameters given above, to derive minimum fluences for each FRB assuming boresight positions in the detection beams. 
Using the 
results summarised in Fig.~5, I also derive 95\% confidence fluence upper limits for each FRB. For FRBs 010724 and 150807, 
I use the constrained $1\sigma$ fluence ranges that were derived above in \S3.2 and by \citet{rsb+16} respectively, enabled by 
their multiple-beam detections. I collate the fluence measurements to derive lower- and upper-limiting empirical logN-logF 
distributions, which are displayed in Fig.~7. 

In the Figure, I also show various logN-logF functions of index $-0.7$, which was the value inferred by \citet{vrh+16}, and of index 
$-3/2$, which corresponds to a uniform distribution of sources in Euclidean space. In 
comparing the empirical distributions with the assumed intrinsic power-law logN-logF functions, a number of selection effects must be 
recognised. First, the data in Fig.~7 are comprised of FRBs detected with both the BPSR and AFB instruments; FRBs detected 
with the AFB are indicated by red arrows. The differing sensitivities of these instruments, attributable to different numbers of 
bits in the analogue-to-digital conversion, the different bandwidths, and different integration times, mean that the instruments are 
fluence-incomplete \citep{kp15} below different thresholds. This threshold is $\approx 2$\,Jy\,ms for BPSR, and 
$\approx 3$\,Jy\,ms for the AFB. Second, it is possible that the FRB rate varies with Galactic latitude \citep{pvj+14,bb14}. 
Certainly, different sky radiation-temperatures result in different sensitivities 
for different pointings.  The combination of FRB detections from varied searches, even with the BPSR instrument, 
may therefore result in a biased estimate of the sky-averaged FRB logN-logF function. 
Given these issues, and factor of $\sim5$ uncertainty in the flux-density values, I concur with \citet{kp15} and \citet{vrh+16} 
that it is not useful to attempt to use the fluence measurements to directly estimate the FRB logN-logF.

\section{Type~I, Type~II ... Type~N FRBs}

In this paper, I have focused on the Parkes FRB sample to attempt to ensure a consistent sample selection in my analysis. 
This is possible in particular with those FRBs detected with the BPSR instrument at Parkes. However, even within the sample of 
FRBs detected with BPSR, a distinction may be made between those FRBs that are consistent with the simple temporal structures 
in my Models 0--3, and the more complex structures seen in FRBs 121002 and 130729. Another distinction may be made between 
those Parkes FRBs that show 
signatures of scattering at levels stronger than expected from their passage through the Milky Way, and those that do not. 
In a broader context, the repeating nature of FRB\,121102 is, prima facie, unique among FRBs. In this section, I explore a 
selection of arguments relating to the number of possible FRB source classes. Unfortunately, it appears difficult at present to 
distinguish between an entirely homogeneous population of FRB sources, FRB sources that are physically similar but 
which vary in their emission properties, and multiple independent populations of FRB sources.  

\subsection{A single FRB population?}

A unified class of FRB emitters must fulfil the following: (\textit{a}) they must be able to emit pulses at frequencies between 700\,MHz \citep{mls+15} and 
5\,GHz \citep{mph+17}, (\textit{b}) they may lie behind plasma regions with either significant or minor scattering strength, and 
(\textit{c}) they must be capable of producing multiple pulses with a variety of morphologies and luminosities that vary by a few 
orders of magnitude. Neutron stars, for example, are capable of producing such a diversity of radio-emission properties, although 
they have not yet been empirically associated with radiation at the luminosities ascribed to FRBs. 

The large quantity of follow-up observations of the Parkes FRB sample \citep{lbm+07,rsj15,pjk+15,rsb+16} implies that 
their characteristic repeat rate is substantially lower than that of FRB\,121102. Twelve FRBs from the Parkes sample have been 
re-observed for a total of 340\,h with identical instrumentation to the detection observations. As no repeat bursts were detected, the 
95\% confidence upper limit on the number of repeat bursts expected in this time is three \citep{g86}. Assuming that repeat bursts within a 
factor of 1.5 in S/N of the original events could have been detected, the 95\% confidence upper limit on the rate of such repeats 
for the Parkes FRBs is $0.009$\,h$^{-1}$. This is substantially smaller than the rate of FRB\,121102 repeats within a factor of 1.5 
of the peak S/N, which, assuming that $3/17$ of the repeats fulfil this criterion \citep{ssh+16,sshc+16}, is $\approx0.06$\,h$^{-1}$. Therefore, if all FRBs were similar to FRB\,121102, this object is an outlier with respect to its repeat rate, or its pulse-fluence
 distribution, or both. 

Additionally, if all FRBs share a common class of emitter with no significant variation in the emission properties across the population, 
the flat FRB logN-logF \citep{cfb+16,vrh+16} suggests that 
the emitter population is not uniformly distributed in the local Universe. Instead, the population may evolve strongly with distance, 
or may be observed at sufficient distance such that cosmological effects become important in relating volume to luminosity distance. 
The substantial redshift of FRB\,121102 \citep{tbc+17}, the localisation of FRB\,150807 \citep{rsb+16}, and the 
large DMs of many of the Parkes FRBs \citep{dgb+15} are all indicative of the cosmological interpretation. 
In this scenario, the Arecibo telescope would be expected to be sensitive to typically more distant FRBs 
than Parkes, given its greater sensitivity. However, FRB\,121102 has a lower redshift than may be inferred for the Parkes FRBs with 
substantially larger DMs, although relating extragalactic DM to distance is significantly uncertain \citep{dgb+15}, 
and this is an argument based on a sample size of unity. 

On the other hand, different FRB properties may be manifested by different sub-types of the same class of emitter. For example, 
young FRB emitters may emit frequent pulses, whereas older FRB emitters may emit pulses only sporadically. Neutron stars, for example, 
exhibit such evolution in their pulse-emission properties. In this case, with respect to modelling of the FRB logN-logF, the older 
emitter population would be considered independent from the younger population, because the intrinsic pulse luminosity-functions 
of the two populations would be different. A larger variety of FRB logN-logF slopes would then be possible, for example if the older 
emitters could be observed at larger (cosmological) distances.

\subsection{Repeating FRBs in the Parkes sample?}

FRB\,121102 is empirically unique among the FRB population, as it is the only known repeating source of extragalactic radio pulses. 
However, the question of whether it is truly unique in its class of emitter can be addressed by considering the following: 
given the relative amounts of time surveyed by Parkes and Arecibo for FRBs, how many more/less repeating FRBs should Parkes 
have detected as compared with Arecibo? Similar analyses have been conducted by 
\citet{sshc+16} and \citet{ocp16}, who assumed that FRB\,121102 and the Parkes FRBs are drawn from the same population, and 
thus found consistency between the detection rates at both telescopes for a uniformly-distributed, Euclidean-space population. 
Here I assume that repeating FRBs are uniformly distributed in Euclidean space, and calculate for each flux-density 
the relative number of detections above that flux density at Parkes and Arecibo. 

I use the Arecibo survey details given by \citet{sshc+16}: 36.9 days searched, with a seven-beam system where each beam has an 
Airy-function response, $B_{\rm AO}(\theta)$, with 3.5\arcmin~FWHM. I also use the Parkes High Time Resolution Universe (HTRU)
survey details given by \citet{cpk+16}: 152 days searched, with a 
13-beam system where each beam has an Airy-function response, $B_{\rm PKS}(\theta)$,  with a 14.4\arcmin~FWHM. To this I 
add the 106 days searched at Parkes by the Survey for Pulsars and Extragalactic Radio Bursts \citep[SUPERB;][]{kbj+17} and its successors (personal communication from E.~Keane). I assume that the Arecibo system is 13.6 times as sensitive as the Parkes system \citep{cpk+16,sshc+16}. 

Note that my assumptions about the beam 
responses are not wholly correct, because the outer beams of both systems probe somewhat larger sky areas. This is, however, a 
negligible factor given other uncertainties about survey locations on the sky relative to Milky Way dispersion and scattering properties, 
radio-frequency interference, and the rejection of multiple-beam detections, all of which I do not include in the analysis. 

For a flux density with arbitrary units given by $B_{\rm PKS}^{-1}(\theta)$, the number of detections above this level at Parkes is proportional to
\begin{align}
\begin{split}
N_{\rm PKS}(\theta) \propto & \int_{0}^{\theta}\sin\theta'B^{3/2}_{\rm PKS}(\theta')d\theta' \\
& \times(258\,{\rm days})\times(13\,{\rm beams}). \end{split}
\end{align}
In the same units, and for a flux density given by $[13.6\times B_{\rm AO}(\theta')]^{-1}$, the number of detections above this level at Arecibo is 
\begin{align}
\begin{split}
N_{\rm AO}(\theta) \propto & \int_{0}^{\theta}\sin\theta'[13.6\times B_{\rm AO}(\theta')]^{3/2}d\theta' \\
& \times(36.9\,{\rm days})\times(7\,{\rm beams}). \end{split}
\end{align}
The inclusion of the effects of the telescope sky-response extends the analyses of \citet{sshc+16} and \citet{ocp16}. 

\begin{figure}
\centering
\includegraphics[scale=0.55]{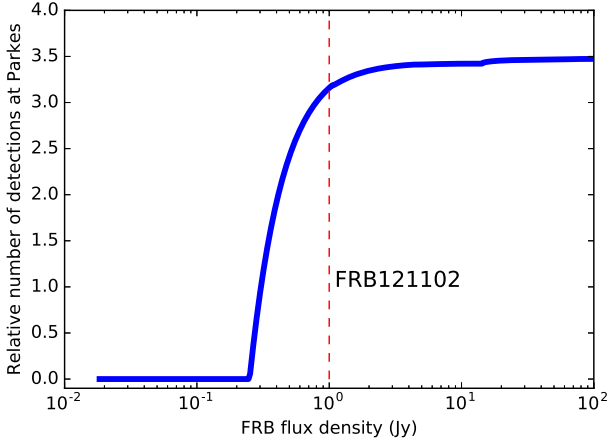}
\caption{Number of detections of objects like FRB\,121102 at Parkes above different flux densities, in the HTRU and SUPERB 
surveys, relative to the number of Arecibo detections. In setting the flux-density scale, I assume FRBs of 3-ms durations, like the 
first-detected pulse from FRB\,121102; the red dashed line indicates the nominal flux density of this pulse, accounting for its 
sidelobe detection.}
\label{fig:9}
\end{figure} 

I plot $N_{\rm PKS}/N_{\rm AO}$ in Fig.~8 for different FRB flux densities, assuming FRB durations of 3\,ms (corresponding 
to the duration of the first-detected pulse from FRB\,121102) and a Parkes MBR system-equivalent flux density of 40\,Jy \citep{kjv+10}. 
For example, 
if FRB\,121102 always emitted 1-Jy pulses of 3-ms duration, like its first-detected pulse accounting for its detection in a 
primary-beam sidelobe, the fact that Arecibo has detected one such object implies that the HTRU and SUPERB surveys 
should expect to have detected three similar objects. If instead FRB\,121102 always emitted pulses of flux-density 0.3\,Jy, Parkes 
should expect to have detected just one such object. 

The expected number of objects like FRB\,121102 in the Parkes surveys, and likely within the existing sample of Parkes FRBs, 
depends on how the characteristic 
L-band flux density (or fluence) of FRB\,121102 is specified. Of the repeat bursts from FRB\,121102 detected at L-band with Arecibo and 
published by \citet{ssh+16} and \citet{sshc+16}, only one (burst 11) likely lies above the Parkes detection threshold. 
It would hence have easily been possible for the first-detected burst from FRB\,121102 to be below the 0.3-Jy threshold in 
Fig.~8 where no Parkes detections are expected given an Arecibo detection. An exact assessment of the characteristic 
flux density or fluence of FRB\,121102 requires a better determination of the distributions of these quantities among its pulses, 
as well as of the statistics of the temporal clustering \citep{ssh+16}. 

The effects of Galactic interstellar scintillation on this analysis are modest. The transition frequency, $\nu_{t}$, between the strong and weak 
scattering regimes for the position of the host of FRB\,121102 is $\nu_{t}=37.9$\,GHz \citep{cl02,clw+17}. As the fractional diffractive 
scintillation bandwidth scales as $(\nu/\nu_{t})^{17/5}$ \citep{w98}, the observing band at Arecibo will contain many scintles; we can hence neglect the effects 
of Galactic diffractive scintillation in the discovery of FRB\,121102. The modulation index due to refractive scintillation is $\sim(\nu/\nu_{t})^{17/30}=0.15$ at 
$\nu=1.4$\,GHz; this is also negligible, in particular given the uncertainty in specifying the characteristic flux density of FRB\,121102 discussed above. At 
Parkes, the HTRU and SUPERB surveys cover the sky outside the Galactic plane \citep[Galactic latitudes $|b|>15$\,deg for the HTRU survey][]{kjv+10,cpk+16} 
in an unbiased sense. Approximately 20\% of the sky has $\nu_{t}\lesssim1.4$\,GHz \citep{w98,cl02}, and the Parkes surveys are therefore generally in the strong scattering 
regime. Nonetheless, the wide fractional bandwidth of the Parkes receiving system ($\sim25\%$), combined with the dramatic dependence of the 
diffractive scintillation bandwidth on frequency, implies that special conditions are required for only a single scintle of an extragalactic source to be present within 
the Parkes band. Refractive scintillation is likely the dominant effect on Parkes FRB detections, with typical modulation indices of $\sim0.5$. As the probability 
density function of refractive-scintillation intensity variations is only mildly skewed \citep{rcb84}, the expectation for the number of objects like FRB\,121102 
present in the Parkes surveys is not likely to be very sensitive to the effects of scintillation. In fact, co-opting the argument of \citet{mj15}, scintillation may cause 
a boost in the number of objects like FRB\,121102 detected at higher Galactic latitudes by Parkes, relative to my analysis. 

Therefore, the best 
interpretation of Fig.~8 that can be presented here is that the HTRU and SUPERB surveys at Parkes could expect to 
contain up to three analogues of FRB\,121102, if these analogues, presumably less distant than FRB\,121102, 
lie above the Parkes detection threshold. The local-Universe location of these analogues suggests that the logN-logF of this population 
is unlikely to be flatter than $F^{-3/2}$, as I have assumed in producing Fig.~8. 
The best candidate analogues of FRB\,121102 among the Parkes sample are clearly 
FRBs 121002, 130729, and possibly 140514. Like the pulses from FRB\,121102, these Parkes FRBs show complex temporal 
structure, and in the cases of FRBs 130729 and 140514, spectral structures concentrated in $\approx100$\,MHz bands. 

\section{Concluding discussion}

I return first to the question of how an FRB may be defined. All FRBs are fundamentally bursts of radio waves which exhibit levels 
of dispersion that exceed predictions for the Milky-Way ionised ISM column density along their specific sightlines. Beyond this, 
FRBs exhibit 
a broad diversity of (dedispersed) durations, scattering signatures, flux densities, and intrinsic temporal and spectral structures. 
In this paper, I have presented an analysis of the individual properties of the sample of seventeen FRBs detected at the Parkes 
telescope with the thirteen-beam L-band receiver. Eight of these FRBs show signatures of scattering at levels significantly 
greater than expected from the Milky Way, with scattering timescales at 1\,GHz ranging between $0.005-32$\,ms. After 
accounting for the scattering, only five Parkes FRBs have pulse widths that are greater than expected from intra-channel 
smearing caused by their dispersions. The fluences of the Parkes FRB sample span a range greater than $0.7-400$\,Jy\,ms. 

My analysis highlights the utility of searching for FRBs with systems that may better resolve the intrinsic pulse durations, because 
a substantial fraction of FRBs ($6/17$ at Parkes) are temporally unresolved. Such systems would require finer filterbank channel 
widths and better time resolutions than currently available, 
which could be achieved using coherent dedispersion techniques or observations at frequencies 
above the L-band. Better temporal resolution will also provide a boost in S/N for short-duration FRBs. Higher-frequency observations 
have the added bonus of being less affected by the temporal broadening observed in $7/17$ Parkes FRBs; the characteristic spectra 
of FRBs are poorly constrained, and the repeating FRB\,121102 has been observed at frequencies up to 5\,GHz. Avoiding the effects 
of scattering may again provide a boost in S/N because of shorter pulse durations, and may indeed provide sensitivity to a population 
of FRBs that are too broad to be detected in the L-band. Conversely, the effects of scattering must be taken into account in 
predicting FRB detection rates for lower-frequency experiments such as the Canadian Hydrogen Intensity Mapping Experiment \citep[CHME; e.g.,][]{nvp+17} and the Hydrogen Intensity and Real-time Analysis Experiment \citep[HIRAX;][]{nbb+16}. 

On the other hand, a search for long-duration FRBs at all frequencies may also be fruitful. Although the number of false candidates in 
single-dish observations increases rapidly with increasing pulse duration \citep{bb10}, and the 
detection S/N decreases as the square-root of the duration for constant fluence, such a search could be carried out by a 
sensitive interferometric system such as those being commissioned for the Jansky Very Large Array.\footnote{https://caseyjlaw.github.io/realfast/} There appears 
to be no reason to expect all FRBs to have the ``millisecond'' duration often quoted in the literature, beyond the 
effects of intra-channel dispersion smearing and scattering, and the increased sensitivity to narrower pulses. 

My revised estimates of FRB scattering timescales, $\tau_{\rm 1\,GHz}$, reveal moderate evidence for a relation between 
$\tau_{\rm 1\,GHz}$ and the extragalactic DM (${\rm DM}_{E}$), similar to that observed for pulsars in the Milky Way 
(Fig.~4). The one outlier is 
FRB\,010724, which has $\tau_{\rm 1\,GHz}=25\pm5$\,ms for a low ${\rm DM}_{E}=288$\,pc\,cm$^{-3}$. The existence of 
such a relation, if supported by further observations, suggests that FRBs are predominantly dispersed in a medium within which 
they are also scattered, and for which the scattering strength increases for larger DM. This medium could be the ISM of 
FRB host galaxies, which would imply modest FRB distances, or the IGM or intervening collapsed systems. Observations of scattering in FRBs with distance 
measurements, obtained for example through localisation and the identification of host galaxies, could resolve the nature of the 
scattering and dispersing medium. 

Although it appears that Parkes FRBs detected in individual beams of the multibeam receiver can have their fluences constrained 
to within a factor of five with 95\% confidence, this is insufficient to directly estimate the FRB flux-density distribution (logN-logF). My 
analysis is therefore unable to distinguish between the case of a uniform distribution of FRB sources in the nearby Universe, 
and the flatter logN-logF distribution expected for a cosmological or evolving FRB population.  

Finally, although the repeating FRB\,121102 is an 
outlier among the FRB population in its repeat rate and the low fluences of most of its bursts, it is not demonstrably unique 
in its class of progenitor. The rate of repeats within the Parkes FRB population is lower than that of FRB\,121102, and 
most Parkes FRBs have simpler temporal and spectral structures. However, some Parkes FRBs are similar in their 
morphologies to bursts from FRB\,121102, and statistical arguments suggest that it is possible that up to three objects like FRB\,121102 have already been detected 
in surveys at Parkes. In a broader context, it is quite possible for all FRBs to be emitted by the same class of astrophysical object, but 
for such objects at different evolutionary stages to emit FRBs with different luminosity functions and rates.

\section*{Acknowledgements}

I thank R. Shannon, H. Vedantham and S. Kulkarni for useful discussions in the course of this work. I also acknowledge the 
efforts of the FRB Catalogue team in maintaining the publicly accessible FRB database. I thank the referee for useful comments that helped 
improve the manuscript.

\appendix

\section{A signal model for FRBs}

In this section, I summarise the signal model for the Parkes FRBs analysed in this paper, and point out specific assumptions that I make. The voltage signal, $V$,  
presented to the AFB and BPSR backends at time $t$ can be represented as follows:
\begin{equation}
V(t) = N(t)+h_{\rm IM}\ast F(t),
\end{equation}
where $N(t)$ is the receiver noise contribution, $h_{\rm IM}(t)$ is a filter that encapsulates the effects of the ISM and IGM on the signal and $F(t)$ is proportional 
to the measured time-varying electric field of the FRB in a single polarisation. I assume that the signal is unpolarised, although this is not particularly relevant to my work. 
I assume that samples of the receiver noise $N(t)$ can be described by a time-stationary 
normal distribution, with zero mean and variance $\sigma_{N}^{2}$ (i.e., $\mathcal{N}(0,\sigma_{N}^{2})$). This assumption neglects the potential effects of RFI. The FRB 
signal can be expressed as $S(t)=A(t)M(t)$, where $A(t)$ is an amplitude envelope, and $M(t)$ is again Gaussian with distribution $\mathcal{N}(0,\sigma_{M}^{2})$. 
I further assume that $\sigma_{M}\ll\sigma_{N}$; that is, I do not account for ``self-noise'' in estimates of FRB properties because FRBs typically contribute negligibly to the 
system temperature \citep[although see][]{rsb+16}. Finally, I note that $V(t)$ is band-limited, and thus correlated on short timescales. 

To illustrate my assumptions for the ISM/IGM effects, consider the Fourier transform of $V(t)$:
\begin{equation}
\tilde{V}(\nu) = \tilde{N}(\nu) + \tilde{h}_{\rm IM}\tilde{S}(\nu),
\end{equation}
where a tilde indicates a frequency-domain quantity. I assume that the ionised-medium filter $\tilde{h}_{\rm IM}$ can be expressed as the product of the standard 
cold, sparse plasma dispersion kernel, $\tilde{h}_{\rm DM}$ \citep{h71}, and the pulse broadening function (PBF) caused by multi-path propagation, $\tilde{h}_{\rm PBF}$. That 
is, 
\begin{equation}
\tilde{h}_{\rm IM} = \tilde{h}_{\rm DM}\tilde{h}_{\rm PBF}.
\end{equation}
The assumption that $\tilde{h}_{\rm DM}$ and $\tilde{h}_{\rm PBF}$ are separable is valid for most, although perhaps not all, pulsars in the Galaxy \citep{css16}. I assume 
a one-sided exponential form for the PBF in the time domain, corresponding to the thin-screen scattering model \citep{c70}, wherein 
\begin{equation}
h_{\rm PBF}(t) = e^{-t/\tau}H(t)n(t),
\end{equation}
where $\tau$ is the scattering timescale, $H(t)$ is the Heaviside step function, and $n(t)$ is a standard-normal random process. 

The AFB and BPSR hardware are used to estimate the signal power at specific frequencies, $\nu_{0}$, within bandwidths $\Delta\nu$ and times $\Delta t$. I model this as follows:
\begin{equation}
\hat{S}(\nu_{0},t) = \int_{t-\Delta t/2}^{t+\Delta t/2} |g(t')\ast V(t')|^{2}dt',
\end{equation}
where $g(t')$ is the time-domain representation of the filter corresponding to a single filterbank channel. I assume the following form for the frequency-domain filter:
\begin{equation}
\tilde{g}(\nu) = \sqrt{\frac{2}{\pi\Delta\nu^{2}}}e^{-2(\nu-\nu_{0})^{2}/\Delta\nu^{2}}.
\end{equation}
That is, I assume that the response of each filterbank channel is a Gaussian function with a standard deviation of $\Delta\nu/2$. Although this model does not accurately 
represent the responses of the analogue filters of the AFB or the polyphase-filterbank channels of BPSR, it appears adequate given the quality of the FRB data. The 
characteristic impulse-response timescale of the estimates of $\hat{S}(\nu_{0},t)$ is therefore $1/\Delta\nu$; for both the AFB and BPSR, $\Delta t\gg1/\Delta\nu$.

\bibliographystyle{mn2e}
\bibliography{mn-jour,vikram}

\end{document}